\documentclass[
reprint,
superscriptaddress,
preprintnumbers,
amsmath,amssymb,
aps,
pra,
floatfix,
]{revtex4-2}
\usepackage[caption=false]{subfig}
\usepackage{algorithm}
\usepackage{algpseudocode}
\usepackage{amsmath}
\usepackage{amssymb}
\usepackage{url}
\usepackage{graphicx}
\usepackage[english]{babel}

\usepackage{dcolumn}
\usepackage{bm}
\usepackage{hyperref}
\usepackage{xcolor}
\usepackage{ifthen}
\newboolean{showcomments}
\setboolean{showcomments}{true}
\ifthenelse{\boolean{showcomments}}
{ \newcommand{\mynote}[3]{
     \fbox{\bfseries\sffamily\scriptsize#1}
        {\small$\blacktriangleright$\textsf{\emph{\color{#3}{#2}}}$\blacktriangleleft$}}}
{ \newcommand{\mynote}[3]{}}

\begin{document}
\title{Spectator-Aware Frequency Allocation in Tunable-Coupler Quantum Architectures}

\author{Evan McKinney}
\email{evm33@pitt.edu}
\affiliation{University of Pittsburgh, Pittsburgh, Pennsylvania, USA}

\author{Israa G. Yusuf}
\affiliation{University of Pittsburgh, Pittsburgh, Pennsylvania, USA}
\affiliation{Yale University, New Haven, Connecticut, USA}

\author{Girgis Falstin}
\affiliation{University of Pittsburgh, Pittsburgh, Pennsylvania, USA}

\author{Gaurav Agarwal}
\affiliation{Yale University, New Haven, Connecticut, USA}

\author{Michael Hatridge}
\affiliation{Yale University, New Haven, Connecticut, USA}

\author{Alex K. Jones}
\email{akj@syr.edu}
\affiliation{Syracuse University, Syracuse, New York, USA}

\begin{abstract}
This paper addresses frequency crowding in SNAIL-based superconducting quantum modules. First, we present design constraints by describing a physical model for realizable gates within a module, and building a fidelity model using error budgeting derived from device characteristics. Second, we tackle the allocation problem by analyzing the impact of frequency crowding on gate fidelity as the radix of the module increases. We explore whether the heuristic gate fidelity can be optimized with a discrete set of qubit frequencies while adhering to defined separation thresholds. By leveraging a combination of analytical and numerical techniques, we demonstrate scalable frequency allocation strategies that minimize spectator-induced errors. Our results further indicate that removing edges leads to improved gate fidelities while maintaining sufficient connectivity, suggesting that edge density is not a limiting factor for NISQ-scale benchmarks. The findings have implications for designing robust, high-fidelity quantum systems with practical constraints on hardware and connectivity.
\end{abstract}

\maketitle

\section{Introduction}

\begin{figure}[t]
    \centering
    \includegraphics[width=\columnwidth]{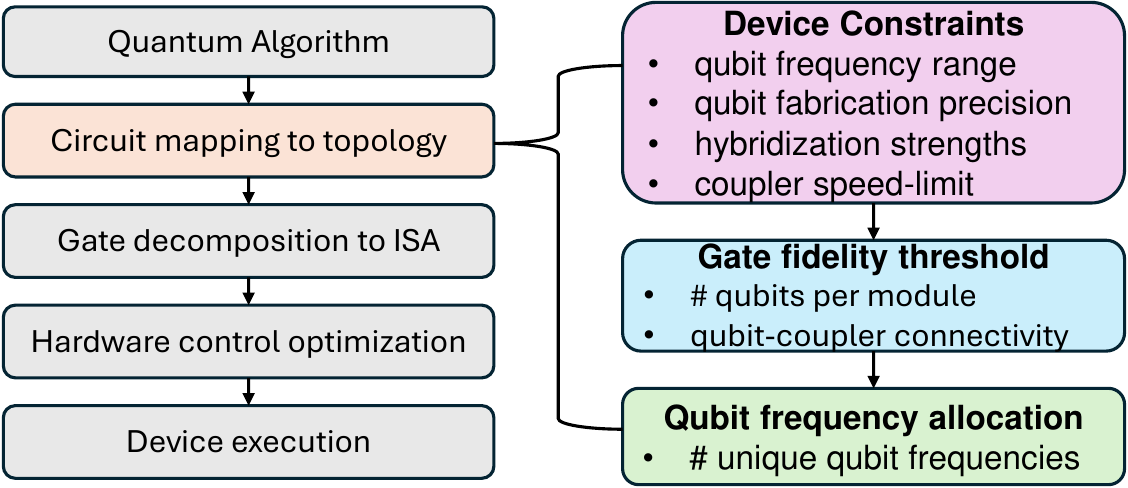}
    \caption{High-level workflow for the co-design of quantum computer architectures~\cite{jones_layered_2012}.}
    \label{fig:codesign-flow}
\end{figure}

Quantum circuits are executed by mapping gates to pulse controls on physically coupled qubit pairs. Accurate execution requires precise control of qubit-qubit couplings, activating interactions only when needed~\cite{beverland_assessing_2022, tomesh_quantum_2021, murali2020software}. However, as coupling density increases, frequency crowding makes it harder to isolate specific interactions, degrading gate fidelity. Still, high connectivity is needed to reduce $\texttt{SWAP}$ overhead and support topological error correction~\cite{murali_full-stack_2019}. Modular superconducting architectures address this by enabling selective, localized coupling within modules.

A central challenge in Noisy Intermediate-Scale Quantum (NISQ) architecture design is balancing qubit density with gate fidelity (Fig.~\ref{fig:codesign-flow}). Sparse layouts reduce crosstalk and yield higher base fidelities, but require more routing overhead. Denser layouts reduce \texttt{SWAP}s but suffer from increased parasitic interactions. We analyze this trade-off by incorporating physical constraints—qubit frequency ranges, coupler bandwidths, and hybridization strengths—into an error budgeting framework to propose architectures meeting fidelity targets under realistic chip constraints~\cite{brink_device_2018, tripathi_operation_2019, ni_superconducting_2023, sete_error_2024}.

Another consideration is coupler-to-qubit ratio. Unlike fixed qubit-qubit links, tunable couplers can mediate multi-qubit interactions. In particular, the Superconducting Nonlinear Asymmetric Inductive eLement (SNAIL) enables all-to-all parametric coupling within a module, at the cost of weak static cross-Kerr terms~\cite{zhou_realizing_2023}. This flexibility expands the architectural design space, provided spectator effects remain controllable.

We study how increasing the number of qubits per SNAIL affects the ability to selectively activate \texttt{iSWAP} gates. While prior work has optimized frequency allocation and fabrication-aware layout~\cite{li_towards_2019, ding_systematic_2020, smith_scaling_2022, morvan_optimizing_2022, osman_mitigation_2023, zhang_qplacer_2024, zhangEfficientFrequencyAllocation2024}, we focus on gate fidelity constraints induced by coupling all module qubits to a single SNAIL using its third-order mixing. Our contributions include:

\begin{itemize}
\item We show that gate fidelity is \textbf{100$\times$ more sensitive to detuning for SNAIL-qubit conversion than for qubit-qubit conversion}, establishing fundamental limits on modular architecture design.
\item We solve a structured frequency allocation problem to determine the largest viable module size under fidelity constraints. We find \textbf{4 qubits per SNAIL} is the practical upper bound for preserving $>0.99$ fidelity.
\end{itemize}

\section{Background}
This section provides a brief overview of two-qubit gates such as \texttt{iSWAP}, introduces the SNAIL-based modular architecture used in our study, and defines the fidelity metrics used to evaluate architectural trade-offs.

\begin{figure}[t]
    \centering
    \includegraphics[width=\columnwidth]{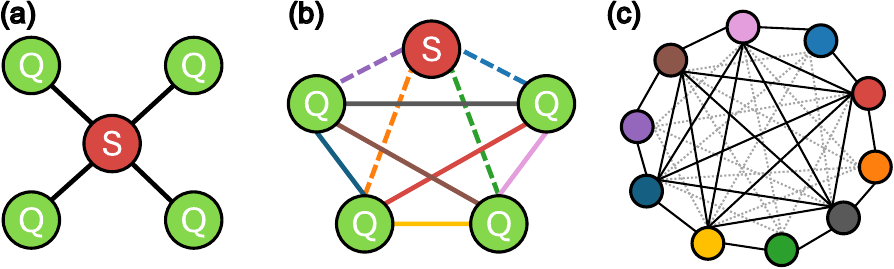}
    \caption{(a) SNAIL-qubit coupling graph (b) 2Q conversion connectivity graph, with unique colored edges for activated exchanges. (c) Compatibility graph, with nodes as interactions and edges as interferences.}
    \label{fig:colored-nodes}
\end{figure}

\begin{figure}[t]
    \includegraphics[width=\columnwidth]{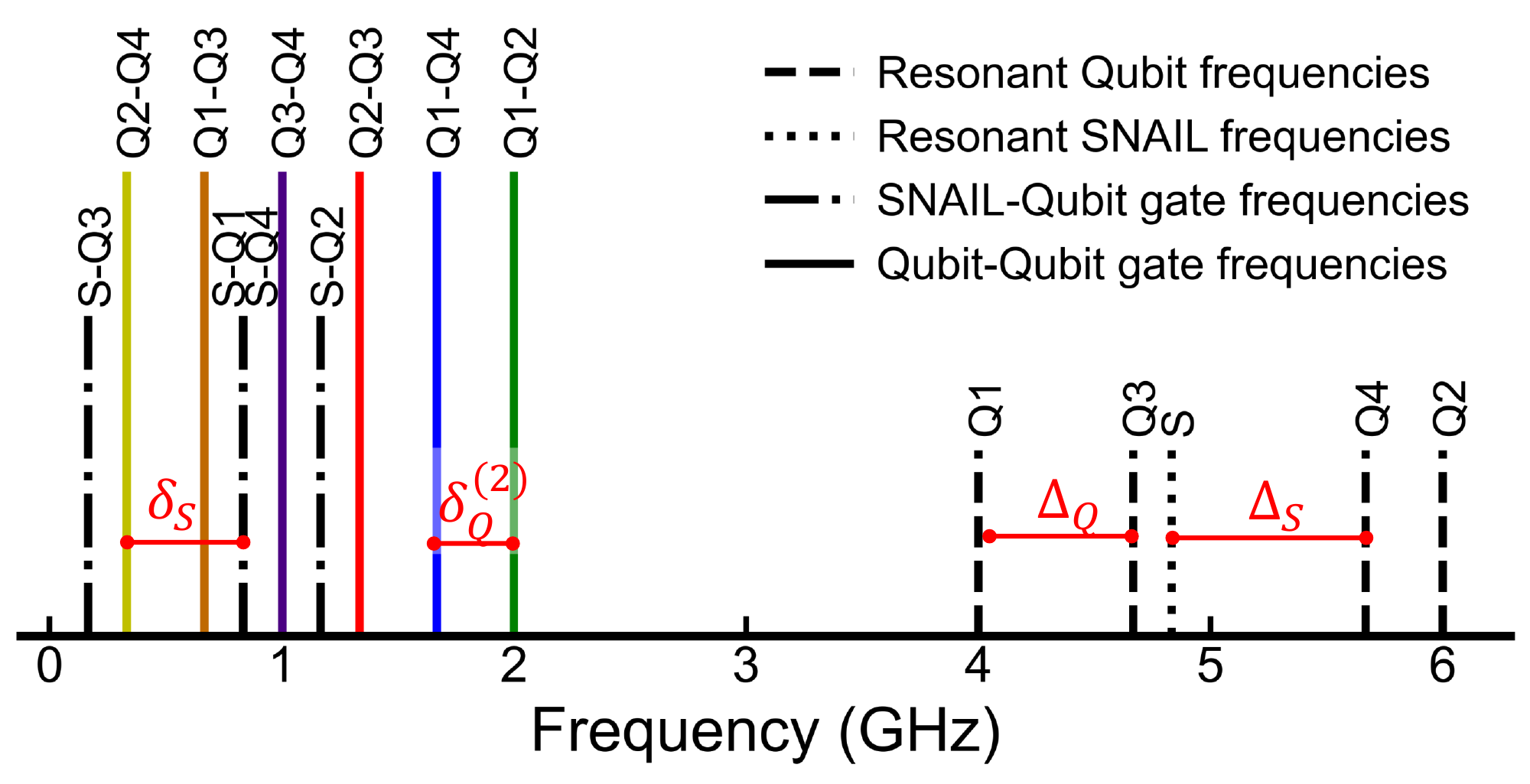}
    \caption{Spectral positioning of a SNAIL and 4 qubit bare modes and their interacting resonant frequencies.}
    \label{fig:spectral_crowding}
\end{figure}


Quantum computation is performed using unitary operations. Single-qubit gates are typically fast and low-error, while two-qubit gates introduce entanglement but are more error-prone. Though \texttt{CNOT} is common in algorithms, the \texttt{iSWAP} gate is more naturally implemented in tunable coupler architectures, enabling direct exchange of quantum states:

\begin{equation}
    \label{iswap}
    \sqrt[n]{\texttt{iSWAP}} =
    \begin{bmatrix}
    1 & 0 & 0 & 0\\
    0 & \cos({\pi/2n}) & \mathit{i}\sin({\pi/2n}) & 0\\
    0 & \mathit{i}\sin({\pi/2n}) & \cos({\pi/2n}) & 0\\
    0 & 0 & 0 & 1
    \end{bmatrix}.
\end{equation}

Gate dynamics are governed by the system Hamiltonian. Assuming time-independent evolution, the unitary is:
\begin{equation}
\label{eq:unitary}
U(t) = e^{-i \int_0^t H(t') dt' / \hbar}  = e^{-i \hat{H} t / \hbar}.
\end{equation} 
Understanding the ideal gate unitary helps identify and mitigate deviations due to parasitic interactions or decoherence.


We implement two-qubit gates using a SNAIL as a tunable coupler~\cite{frattini17}. The SNAIL allows selective activation of specific interaction terms via targeted parametric pumping. Each SNAIL ``Corral" module~\cite{zhou_realizing_2023, mckinney_co-designed_2023} contains four fixed-frequency transmons, each with a single Josephson junction (JJ) that provides dominant fourth-order nonlinearity~\cite{Koch07}. The SNAIL has an asymmetric loop of JJs, which when flux-biased to a critical value creates a device with strong third-order nonlinearity \cite{frattini17}. Each qubit is capacitvely coupled to a readout resonator and controlled via dedicated drive lines.

Parametric drives at resonant conversion frequencies induce hybridization-mediated exchange between qubit pairs. Gate performance thus depends on carefully managing interaction frequencies to avoid unintended exchanges and spectator errors (Fig.~\ref{fig:colored-nodes}).


Quantum circuits incur both coherent (unitary) and incoherent (dissipative) errors. Coherent errors stem from unwanted Hamiltonian terms—e.g., crosstalk and spectator interactions. Incoherent errors include energy relaxation ($T_1$) and dephasing ($T_\phi$), which irreversibly degrade state fidelity. 
We define fidelity heuristically to capture both error types. The average gate infidelity quantifies deviation from the target unitary~\cite{nielsen_simple_2002}:

\begin{equation} 
\label{eq:avg_fidelity}
\epsilon_{\text{avg}} = 1 - \frac{d + \text{Tr}[U V^\dagger]}{d +1},
\end{equation}

where $U$ is the ideal gate, $V$ is the noisy implementation, and $d$ is the Hilbert space dimension. Assuming additive contributions from spectator effects, total coherent infidelity is:
\begin{equation}
\epsilon_{\text{coh}} \approx \sum_{i} \epsilon_\text{avg}^{(i)},
\end{equation}
where $\epsilon_\text{avg}^{(i)}$ denoting infidelity from the $i$th spectator. We model incoherent decay as an exponential suppression: $\epsilon_\text{inc} \approx 1 - e^{-t/\tau}$. The combined fidelity is:
\begin{equation}
\label{eq:total_fidelity}
\epsilon_{\text{gate}} \approx 1 - (1-\epsilon_\text{inc}) \times (1 - \epsilon_{\text{coh}}).
\end{equation}
This decomposition isolates unitary and non-unitary effects. Spectator-induced infidelities add, while decoherence applies a multiplicative penalty. This model underlies our cost function for frequency allocation (Sec.\ref{sec:allocation})~\cite{hopf2025improving, gokhale2024faster, schmid2024computational}.
\section{Infidelity in a Module}

In this section, we characterize how different types of infidelity contribute to errors within a module. We first quantify infidelity per detuning for various interactions. The dominant error sources are spectator terms, which accumulate as coherent infidelity. Additionally, subharmonic constraints of the SNAIL coupler impose limits on gate speed, contributing to incoherent infidelity. These effects together establish the frequency separation constraints used later in chip-level allocation.

\begin{figure}[t] 
    \centering 
    \includegraphics[width=\columnwidth]{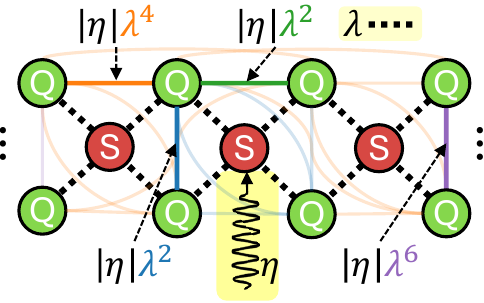} 
    \caption{In this diagram, the central SNAIL is driven with the target interaction denoted by a green edge. Spectator terms diminish based on how many orders of hybridization they are removed from the driven SNAIL. (Blue) Both qubits coupled to the driven SNAIL (in the same module); (Orange) One qubit directly coupled to the driven SNAIL; (Purple) Neither qubit directly coupled to the driven SNAIL.} 
    \label{fig:corral}
\end{figure}

\subsection{Hamiltonian Expansion}
To derive the two-qubit basis gates, we begin with a static system Hamiltonian and apply standard transformations to move into the interaction frame, allowing each term to acquire a pump-frequency-dependent coefficient.

The SNAIL-Corral architecture implements an \texttt{iSWAP} instruction set~\cite{chen_one_2023} using photon-conversion interactions. The full Hamiltonian is decomposed as $H = H_{0L} + H_{NL} + H_c$, with $H_{0L}$ capturing the bare mode frequencies:

\begin{equation}
H_{0L} = \omega_s s^\dagger s + \sum_{i} \omega_i q_i^\dagger q_i,    
\end{equation}

where $\omega_s$ and $\omega_i$ denote the SNAIL and qubit frequencies. The nonlinear terms are:
\begin{equation}
H_{NL} = g_3 (s^\dagger + s)^3 + \sum_{i} \frac{\alpha_i}{12}(q_i^\dagger + q_i)^4,
\end{equation}
with $g_3$ the third-order SNAIL nonlinearity and $\alpha_i$ the transmon anharmonicity. Coupling between SNAIL and qubits is defined by:
\begin{equation}
H_c = \sum_q g_{sq} (s^\dagger q + s q^\dagger),
\end{equation}
where $g_{sq}$ is the coupling strength.

The interaction frame is useful for describing an effective Hamiltonian where each interaction term is modulated by a rate (the strength of the pump) and a complex phase factor (how far off-resonance the interaction is from the pump's frequency). Transforming to the interaction frame via the Bogoliubov diagonalization and displacement transform~\cite{zhou_superconducting_2023,zhou_realizing_2023, xia2023fast}, the Hamiltonian becomes:
\begin{multline}
\label{eq:effective-hamiltonian}
\tilde{H}_I = g_3 \left( s e^{-i \tilde{\omega}_s t} + \eta e^{-i \omega_p t} + \sum_{i} \lambda_{si} q_i e^{-i \tilde{\omega}_{q_i} t} + \text{h.c.} \right)^3 \\
\quad + \sum_{i} \frac{\alpha_i}{12} \left( q_i e^{-i \tilde{\omega}_{q_i} t} - \lambda_{si}(s e^{-i \tilde{\omega}_s t} + \eta e^{-i \omega_p t}) + \text{h.c.} \right)^4.\end{multline}

Here, $\tilde{\omega}$ are dressed mode frequencies, and $\lambda_{si} = g_{sq}/\Delta$ quantifies hybridization between the SNAIL and each qubit. The pump strength $\eta = \sqrt{n_s}$ reflects the coherent pump amplitude in the SNAIL mode. Terms are resonant when $\tilde{\omega}_x \approx \omega_p$, and vanish when far-detuned. The Rotating-Wave Approximation (RWA) retains only those terms near resonance, this \textit{assumes that the unwanted terms are actually far enough off-resonance} and thus could be considered sufficiently small~\cite{zhou_superconducting_2023,zhou_realizing_2023, xia2023fast,barajas2025quantum}.

We use OpenFermion and SymPy to symbolically expand the cubic and quartic Hamiltonian polynomials (Eq.~\ref{eq:effective-hamiltonian}), in addition to normal-ordering and expression simplification. We retain dominant terms: two-qubit \texttt{iSWAP}s, SNAIL-qubit conversions, and subharmonic single-mode drives (Table~\ref{tab:order_magnitude_terms}). Not included, \textit{e.g.}, are gain-process terms, conjugate operators created by frequency additions rather than differences, because they are far from the pump.

\begin{table}[t]
\centering
\caption{Order of magnitude analysis for driven, intra-module, and inter-module spectator terms sorted by normalized prefactor. Here, $q_a$ and $q_b$ are generic qubits in the driven module; $s$ denotes the SNAIL; and $q_c$, $q_d$, $s_n$ denote neighboring module elements.}
\resizebox{\columnwidth}{!}{
\begin{tabular}{|c|c|c|c|}
\hline
\multicolumn{4}{|c|}{\textbf{Driven Term}} \\ \hline
Term & Coefficient & $\omega_p =$ & Normalized Prefactor \\ \hline
$(q_a^\dagger q_b + q_a q_b^\dagger)$ & $6 |\eta| \lambda^2 g_3$ & $|\omega_{q_b} - \omega_{q_a}|$ & 1.0 \\ \hline
\multicolumn{4}{|c|}{\textbf{Intra-Module Spectator Terms}} \\ \hline
$(s^\dagger + s)$ & $3 |\eta|^2 g_3$ & $\omega_s / 2$ & 100.0 \\ \hline
$(s^\dagger q_a + s q_a^\dagger)$ & $6 |\eta| \lambda g_3$ & $|\omega_s - \omega_{q_a}|$ & 10.0 \\ \hline
$(q_a^\dagger + q_a)$ & $3 |\eta|^2 \lambda g_3$ & $\omega_{q_a} / 2$ & 10.0 \\ \hline
$(s^\dagger q_a + s q_a^\dagger)$ & $\alpha |\eta|^2 \lambda^3$ & $|\omega_s - \omega_{q_a}| / 2$ & 0.067 \\ \hline
$(q_a^\dagger + q_a)$ & $\alpha |\eta|^3 \lambda^3/3$ & $\omega_{q_a} / 3$ & 0.044 \\ \hline
$(s^\dagger + s)$ & $N_q \alpha |\eta|^3 \lambda^4/3$ & $\omega_s / 3$ & 0.018 \\ \hline
\multicolumn{4}{|c|}{\textbf{Inter-Module Spectator Terms}} \\ \hline
$(s_n^\dagger + s_n)$ & $3 |\eta|^2 \lambda^2 g_3$ & $\omega_{s_n} / 2$ & 1.0 \\ \hline
$(s^\dagger q_c + s q_c^\dagger)$ & $6 |\eta| \lambda^3 g_3$ & $|\omega_s - \omega_{q_c}|$ & 0.1 \\ \hline
$(q_c^\dagger + q_c)$ & $3 |\eta|^2 \lambda^3 g_3$ & $\omega_{q_c} / 2$ & 0.1 \\ \hline
$(q_a^\dagger q_c + q_a q_c^\dagger)$ & $6 |\eta| \lambda^4 g_3$ & $|\omega_{q_c} - \omega_{q_a}|$ & 0.01 \\ \hline
$(s_n^\dagger q_a + s_n q_a^\dagger)$ & $6 |\eta| \lambda^5 g_3$ & $|\omega_{s_n} - \omega_{q_a}|$ & 0.001 \\ \hline
$(q_c^\dagger q_d + q_c q_d^\dagger)$ & $6 |\eta| \lambda^6 g_3$ & $|\omega_{q_d} - \omega_{q_c}|$ & $0.0001$ \\ \hline
\end{tabular}
}
\label{tab:order_magnitude_terms}
\end{table}

To isolate a desired two-qubit gate, we set the pump on-resonance, $\omega_p = \tilde{\omega}{q_2} - \tilde{\omega}{q_1}$, selecting the term:
\begin{equation}
H_{\text{target}} = 6 |\eta| g_3 \lambda^2 (q_1 q_2^\dagger + q_1^\dagger q_2).
\end{equation}
The required pulse duration $t_f$ follows from Eq.~\ref{eq:unitary}, yielding:
\begin{equation}
\label{eq:iswap-eta}
\frac{\pi}{2n} = 6 t_f |\eta| g_3 \lambda^2 \quad \text{(}\sqrt[n]{\text{iSWAP}}\text{)}.
\end{equation}

Spectator interactions result from residual hybridization between qubits and the SNAIL. We distinguish intra-modular spectators—where the qubit is directly coupled to the driven SNAIL—from inter-modular ones, which involve indirect pathways through neighboring SNAILs (Fig.\ref{fig:corral}). The strength and detuning of these interactions determine their contribution to infidelity, as detailed in Table\ref{tab:order_magnitude_terms}.

\subsection{Spectator Infidelities}
\label{sec:simulation-constraints}
To establish frequency separation constraints, we evaluate how detuning impacts gate fidelity. Coherent errors arise from unintended spectator interactions, while incoherent errors are linked to SNAIL coupler speed limits.

\subsubsection{Coherent loss from spectators}
\begin{figure}
    \centering
    \includegraphics[width=\columnwidth]{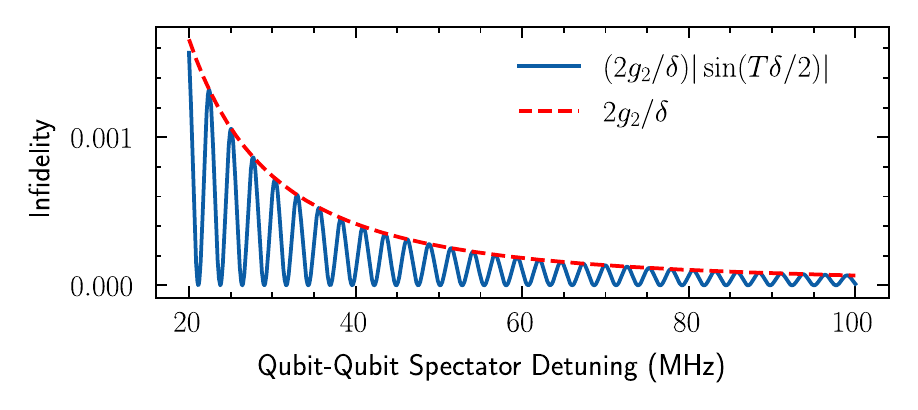}
    \caption{Infidelity versus detuning for spectator amplitude with Rabi oscillations compared with spectator amplitude using Rabi magnitude bound.}
    \label{fig:detuning_rabi}
\end{figure}

To quantify spectator-induced infidelity, we isolate each interaction term and simulate its impact using Eq.~\ref{eq:avg_fidelity}. A representative Hamiltonian with one target and one detuned spectator gate is:
\begin{equation} 
H(t) = 6 |\eta| g_3 \lambda^2 \left( q_1^\dagger q_2 + e^{-it \delta_Q} q_3^\dagger q_4 + \text{h.c.} \right).
\end{equation}

Here, $\delta_Q$ denotes the detuning between the spectator term and the target gate.

In certain cases, the target gate may coincide with the anti-node of the spectator oscillation (Fig.~\ref{fig:detuning_rabi}), partially mitigating its effect. However, in realistic devices where interaction rates vary and gate durations are not as precisely synchronized nor fabricated. Instead, we apply a conservative amplitude bound:
\begin{equation}
\left| \int_0^{T} e^{-i t \delta} dt \right| = 2 | \sin (T \delta /2) |  / \delta \leq 2/\delta.
\end{equation}

This avoids dependence on precise Rabi synchronization and simplifies the simulation by removing time dependence from the interaction term. The resulting unitary becomes:
\begin{equation}
\label{eq:spectator}
U(t) = e^{-i \left (g_1 t (q_1^\dagger q_2) + 2 g_2  (q_3^\dagger q_4)  / \delta  + h.c. \right)},
\end{equation}

with $g_1$ and $g_2$ determined by the target and spectator coefficients in Table\ref{tab:order_magnitude_terms}. While the example above uses a spectator qubit-qubit term $q_3^\dagger q_4 + q_3 q_4^\dagger$, we apply the same procedure to other off-resonant interactions in the same way. The corresponding infidelity is then computed across a range of $\delta_Q$ values and fit to:
\begin{equation}
\label{eq:coh_cost}
\epsilon_\text{coh}(\delta) := \frac{2 x_0}{(x_1 + \delta)^2}  
\end{equation}  
as shown in Fig.~\ref{fig:fidelity_vs_terms}.

\subsubsection{Incoherent loss from speed limits}
\label{sec:snail-death}

\begin{figure}[t]
    \centering
    \includegraphics[width=\columnwidth]{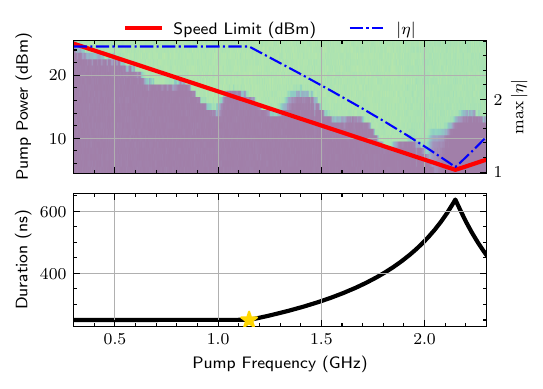}
    \caption{(a) Maximum pump power vs. pump frequency. Overlaid is the maximum achievable $|\eta|$, fitted under the assumption that a viable gate duration exists at the star marker. (b) Gate duration vs. pump frequency, using the star marker to establish a scaling factor between pump power and $|\eta|$.}
    \label{fig:snail_limits}
\end{figure}

We next consider incoherent loss from pump power constraints. Figure~\ref{fig:snail_limits} shows experimental SNAIL breakdown data~\cite{zhou_superconducting_2023, mckinney2023parallel}, where exceeding a pump threshold causes loss of parametric behavior. Although the detailed physics are complex, we fit this data to approximate the maximal achievable $\eta$ as a function of pump frequency.

Driving the SNAIL resonantly enables gate operations by populating it with photons. However, excessive pump power causes breakdown, particularly in subharmonic modes that are $\sim$100$\times$ stronger than typical interaction terms (Table~\ref{tab:order_magnitude_terms}). While these do not directly introduce unitary error, they destabilize gate execution and affect layout density and parallelization~\cite{dumas2024unified}.

From Eq.~\ref{eq:iswap-eta}, gate duration $t_f$ is determined by $\eta$, which itself depends on $\omega_p$ and $\omega_s$:
\begin{equation}
|\eta| = \frac{\epsilon \omega_s}{\omega_p^2 - \omega_s^2},
\end{equation}
Maintaining fixed gate time at low $\omega_p$ requires increasing drive amplitude $\varepsilon$, trading off gate speed against stability.

To model this tradeoff, we assume a linear relationship between dBm and $\varepsilon$, justified by attenuation being roughly linear across the drive chain. This allows us to define a breakdown threshold where increasing $\varepsilon$ further would destabilize the coupler. Above this threshold, additional speedup is infeasible. The critical photon number~\cite{frattini2021three} limits how strongly the SNAIL can be driven. Beyond this, nonlinear effects and chaotic behavior disrupt gate performance~\cite{xia2023fast}. Other second-order effects—such as pump-induced frequency shifts and higher-order Kerr terms—further limit fidelity but are not modeled explicitly here~\cite{Raman2018}.

\begin{figure}[t]
    \centering
    \includegraphics[width=\columnwidth]{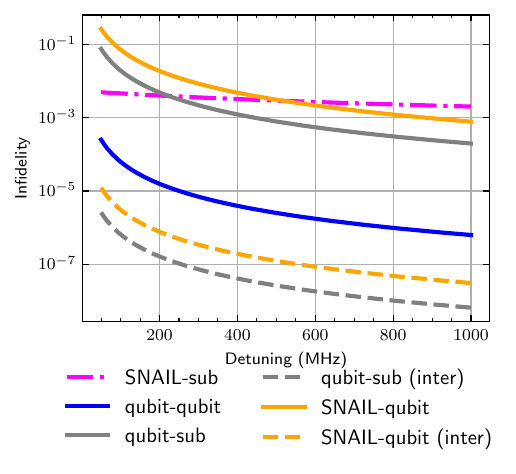}
    \caption{Infidelity scaling versus detuning for different spectator types.}
    \label{fig:fidelity_vs_terms}
\end{figure}

Assuming a viable gate exists at the star marker in Fig.\ref{fig:snail_limits} (250 $\mu$s at 1 GHz detuned from $\omega_S/2$), we fit the maximum usable $\eta$ curve and compute $t_f$. Using Eq.\ref{eq:total_fidelity}, we convert gate duration into decoherence-induced loss and fit the result with the heuristic:
\begin{equation}
\label{eq:incoh_cost}
\epsilon_\text{inc}(\delta) := \frac{x_0}{(x_1 + \delta)}
\end{equation}
which is plotted in Fig.~\ref{fig:fidelity_vs_terms}.

\subsection{Design Tradeoffs}
\label{sec:tradeoffs}
\begin{figure}[t]
    \centering
    \includegraphics[width=\columnwidth]{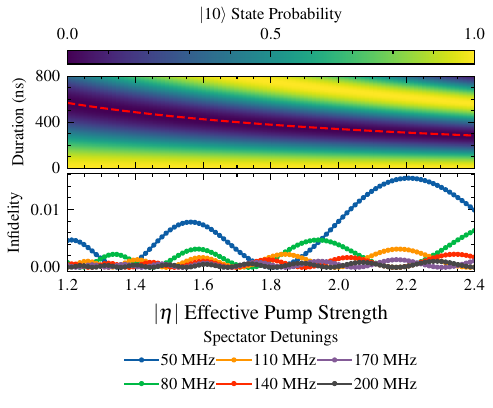}
    \caption{(a) Population exchange between $|01\rangle$ and $|10\rangle$ as a function of time and pump strength. The dashed red line marks a full \texttt{iSWAP}. (b) Coherent infidelity from a qubit-qubit spectator vs. effective pump strength at various $\delta_Q$.}
    \label{fig:infidelity_vs_eta}
\end{figure}

Optimizing gate fidelity requires balancing gate speed against spectator interactions. Stronger pumps shorten gate time but amplify unwanted interactions (Fig.~\ref{fig:infidelity_vs_eta}). From Eq.~\ref{eq:spectator}, achieving $g_1 t_f \gg 2g_2 / \delta_Q$ ensures the target gate dominates. Increasing $\delta_Q$ suppresses spectators but lengthens $t_f$, increasing $T_1$-induced loss.

\begin{figure}[t]
    \centering
    \centering
    \includegraphics[width=\columnwidth]{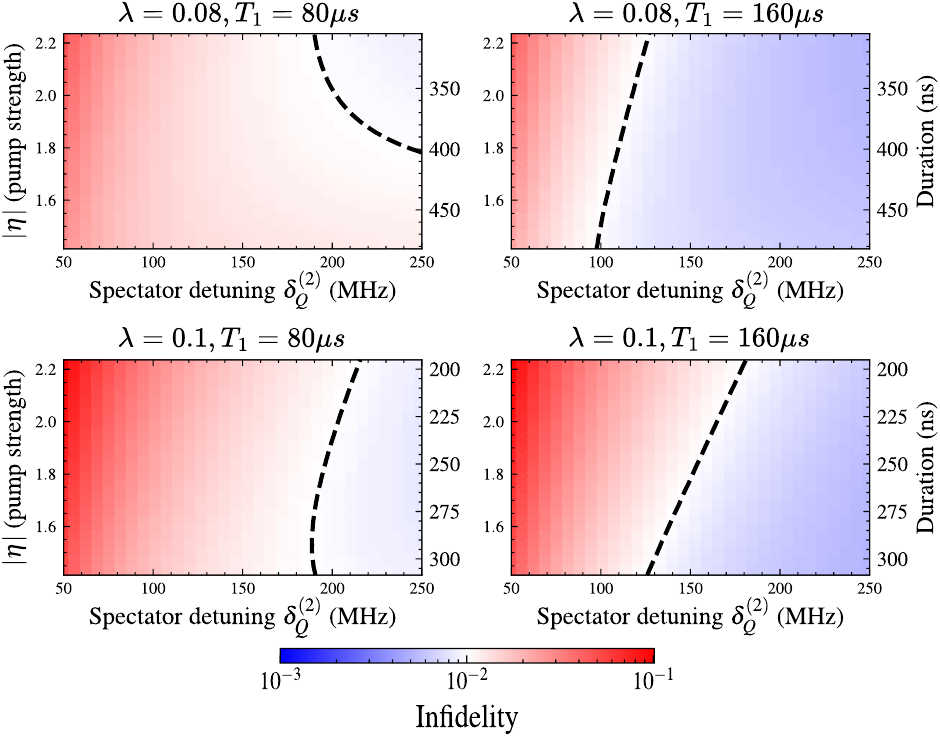}
    \caption{Comparison of \texttt{iSWAP} fidelity threshold boundaries for different values of $\lambda$ and $T_1$.}
    \label{fig:combined-results}
\end{figure}
To evaluate both error types, we simulate noisy evolution using QuTiP's Lindblad solver with amplitude damping for $T_1$. The resulting operator $U(t_f)$ defines the fidelity threshold region in Fig.~\ref{fig:combined-results}, plotted over $\lambda$ and $T_1$. High-fidelity gates ($F \geq 0.99$) appear after around 150–200 MHz detuning. Weaker couplings shift the threshold left, into $T_1$-dominated loss; stronger pumps shift it upward, requiring more detuning to suppress spectators. This reveals an optimal operating region where both error sources are balanced—supporting the bounds in Fig.~\ref{fig:snail_limits}, which limit $\eta$ accordingly.

We assume uniform SNAIL-qubit coupling ($\lambda_i \equiv \lambda_j$), though fabrication variability causes nonuniform hybridization. This affects both gate duration and spectator interference (Fig.~\ref{fig:detuning_rabi}). We also neglect higher-order nonlinearities in the transmon potential, which can cause AC Stark shifts and access higher excitation levels. Finally, we ignore direct qubit-qubit and SNAIL-SNAIL couplings ($\lambda_{qq} = \lambda_{ss} = 0$), though these may be non-negligible when modules are densely packed. In particular, couplers sharing qubits can introduce residual cross-coupling due to their physical proximity, complicating interaction isolation.

\section{Qubit Frequency Allocation}
\label{sec:allocation}

\begin{figure}[t]
    \centering
    \includegraphics[width=\columnwidth]{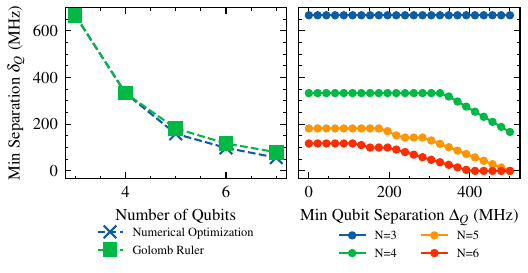}
    \caption{(Left) Minimum Difference Separation vs. Number of Qubits. Numerical optimization scales with the analytical Golomb Ruler but trails the optimal solution due to slight instability in convergence. (Right) Separation results with additional constraint of minimum allowed spacing between bare qubit frequencies.}
    \label{fig:interaction_separation_combined}
\end{figure}

We now use the infidelity models from Section~\ref{sec:simulation-constraints} to construct frequency allocations that maximize gate fidelity. We formalize the frequency allocation task as a variant of the classical Frequency Assignment Problem (FAP), describe our numerical optimization strategy, and evaluate trade-offs that emerge in multi-module layouts.

FAP is a well-known NP-complete problem in which discrete frequency channels are assigned to transmitters while avoiding interference~\cite{park_application_1996, waters2005graph, orden_spectrum_2018}. It is typically cast as a graph coloring problem: nodes are transmitters, colors are frequencies, and edges encode minimum required spacing:
\begin{equation} 
|f(i) - f(j)| \geq q_{ij} \quad \forall (i, j), i \neq j 
\end{equation}

While the classical objective is often to minimize total bandwidth, alternate formulations maximize the minimum pairwise spacing::
\begin{equation} 
\max \min_{i, j \in X, i \neq j} |f(i) - f(j)| 
\end{equation}

Quantum frequency allocation differs fundamentally from classical FAP in that physical constraints act on differences between frequencies, not their absolute values. Two-qubit gates are mediated by frequency detunings, and nonlinearities like subharmonic and hybridization effects impose coupled constraints across all interactions. These dependencies create a rugged optimization landscape that cannot be decomposed into independent frequency assignments.


To quantify achievable interaction separation, we establish a theoretical baseline using a Golomb ruler as an optimal bound. A Golomb ruler is a sequence of marks (interpreted here as frequencies) where all pairwise differences are unique, minimizing spectral crowding~\cite{memarsadeghi2016nasa}. One construction is given by:
\begin{equation}
f_k = f_\text{min} + c (2pk + (k^2 \bmod p)), \quad k = 0, 1, \dots, p-1
\end{equation}
where $p$ is the number of qubits and $c$ a scaling factor. Though suspected NP-hard~\cite{meyer2009complexity}, optimal sequences are known for small $p$, providing a benchmark for interaction separation.

However, the Golomb sequence requires marks to become increasingly close together, focusing only on relative differences while ignoring absolute placement constraints. In quantum systems, overly-close qubit frequencies create new challenges; for example, if qubits are too close in frequency, single-qubit gates become difficult to address, and higher-order state transitions introduce additional interference. To account for this, we introduce a constraint $\Delta_Q$ on the minimum allowed spacing between bare qubit frequencies.

\begin{algorithm}[H]
\caption{Minimize Spectator Infidelity}
\label{alg:numerical_optimizer}
\begin{algorithmic}[1]
\Require Module graph, frequency bounds $\omega_Q, \omega_S$,$\epsilon_{\text{coh}}, \epsilon_{\text{inc}}$, worst-gate exclusion $k$
\Ensure Optimized frequency assignment minimizing total infidelity

\State Initialize random frequencies $\{ \omega_{q_i} \}, \omega_S$
\Repeat
    \State Collect all spectator frequencies $\mathcal{S} = \{ \omega_{\text{spec}} \}$ for all SNAIL-qubit, qubit-qubit pairs and qubit subharmonics
    \For{each gate $(q_i, q_j)$ in module}
        \State $\epsilon_{\text{coh}}(q_i, q_j) \gets 0$
        \For{each spectator $\omega_{\text{spec}} \in \mathcal{S}$}
            \State $\epsilon_{\text{coh}}(q_i, q_j) \mathrel{+}= \epsilon_{\text{coh}}^{\text{spec}}(|\omega_{q_i} - \omega_{\text{spec}}|)$
        \EndFor
        \State $\epsilon_{\text{inc}}(q_i) \gets \epsilon_{\text{inc}}(|\omega_{q_i} - \omega_S / 2|)$
        \State $\epsilon_{\text{gate}}(q_i, q_j) \gets 1 - (1 - \epsilon_{\text{coh}}(q_i, q_j))(1 - \epsilon_{\text{inc}}(q_i))$
    \EndFor
    \For{each qubit}
        \State Apply penalty if $\min |\omega_{q_i} - \omega_{q_j}| < \Delta_Q$
    \EndFor
    \State Sort $\{ \epsilon_{\text{gate}} \}$ in descending order and drop worst $k$ gates
    \State $\mathcal{L} \gets \sum_{\text{gates remaining}} \epsilon_{\text{gate}}$
    \State Update $\{ \omega_{q_i} \}, \omega_S$ using Nelder-Mead to minimize $\mathcal{L}$
\Until{convergence or max iterations reached}
\State \Return Optimized $\{ \omega_{q_i} \}, \omega_S$
\end{algorithmic}
\end{algorithm}

Figure~\ref{fig:interaction_separation_combined} compares our results to the Golomb bound. For small modules, the numerical solutions generally match the ideal separation. When $\Delta_Q$ is enforced, the Golomb bound itself begins to degrade. With four qubits, the optimizer maintains $>300$ MHz spacing; with five, the worst pair drops below 200 MHz. This highlights the challenge of balancing single-qubit isolation against detuning for two-qubit gates.

Unlike the linear FAP, our cost function is nonlinear and involves absolute values, detuning-dependent infidelity models, and coupled penalty terms. We use a numerical optimizer (Algorithm~\ref{alg:numerical_optimizer}) to assign qubit and coupler frequencies that minimize the total two-qubit gate infidelity, incorporating both coherent (spectator) and incoherent (lifetime) loss terms. Related methods have also been explored using graph neural networks~\cite{ai2024graph} and perturbative corrections~\cite{mammola2025optimal}.

The scaling of spectator effects is derived from order-of-magnitude estimates in Table~\ref{tab:order_magnitude_terms}, which categorize the impact of different interaction types. Figure~\ref{fig:fidelity_vs_terms} translates these estimates into infidelity scaling, allowing the optimizer to focus on mitigating the most significant error sources. The largest contributors are coupler-qubit and qubit subharmonic, followed by qubit-qubit interactions. The full optimization process minimizes total infidelity for each gate:
\begin{equation} 
\epsilon_{\text{gate}}(q_i, q_j) = 1 - (1 - \epsilon_{\text{coh}}(q_i, q_j))(1 - \epsilon_{\text{inc}}(q_i)). 
\end{equation}

under the following constraints:
\begin{align}
&3.3 \text{ GHz} \leq \omega_Q \leq 5.7 \text{ GHz}\\
&4.2 \text{ GHz} \leq \omega_S \leq 4.7 \text{ GHz}\\
&|\omega_{q_i} - \omega_{q_j}| \geq \Delta_Q = 200\text{ MHz}.
\end{align}

\begin{figure}[t]
    \centering 
    \includegraphics[width=\columnwidth]{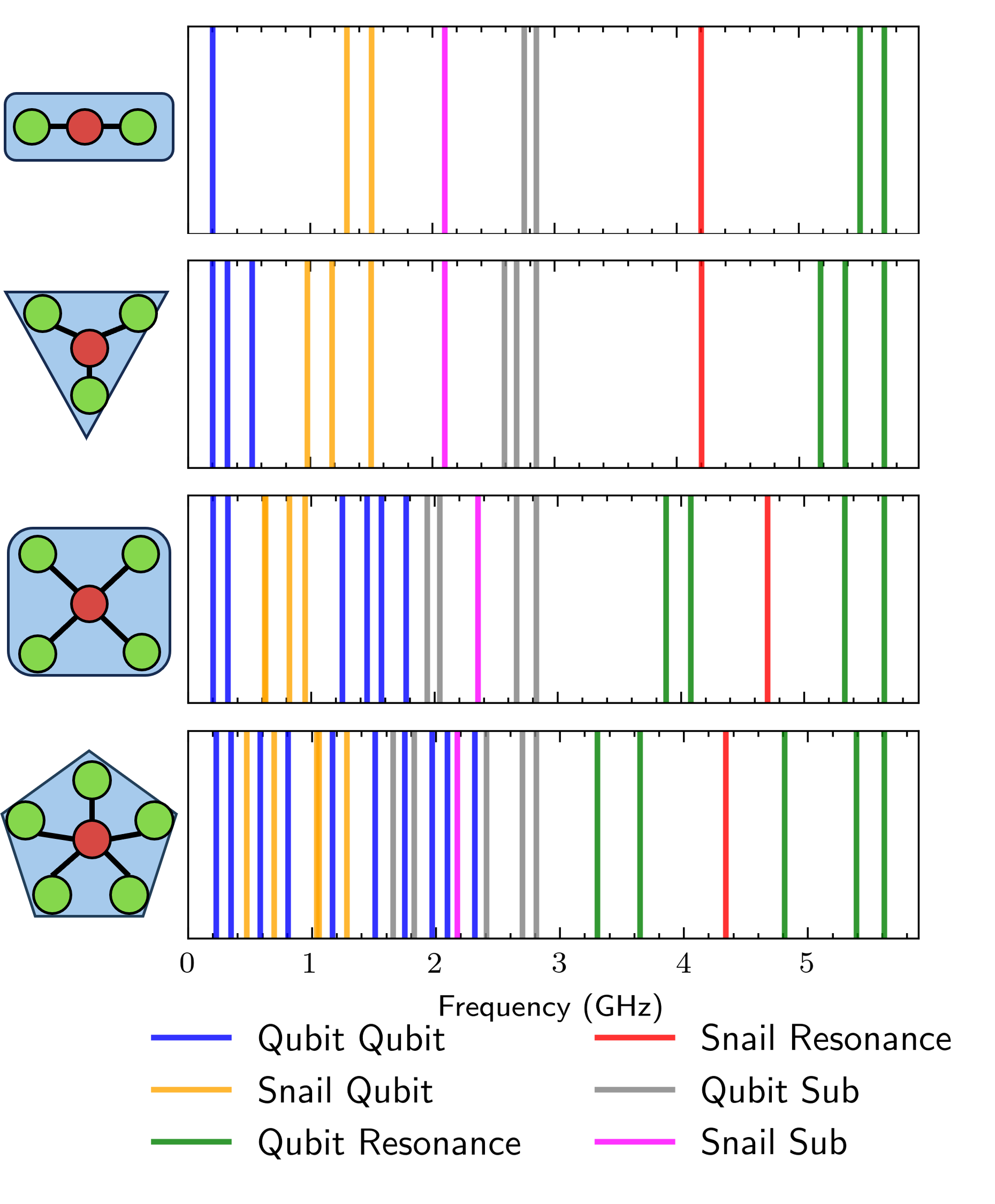}
    \caption{Optimized Frequency Stack: Frequencies of qubit and SNAIL resonances alongside interaction terms, grouped by module size $N = 2, 3, 4, 5$ (from top to bottom).}
    \label{fig:frequency_stack}
\end{figure}

\begin{figure}[t]
    \centering
    \includegraphics[width=\columnwidth]{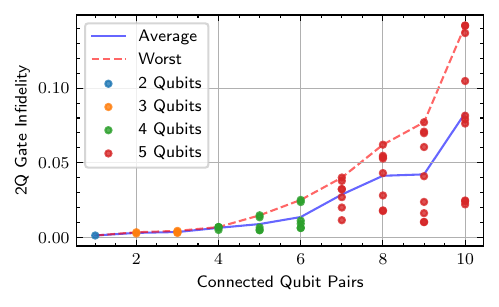}
    \caption{Average two-qubit gate infidelities across module sizes with and without selective edge removal.}
    \label{fig:base-fidelities}
\end{figure}

The optimization proceeds via Nelder-Mead, with heavy penalties imposed for violations of $\Delta_Q$. Figure~\ref{fig:frequency_stack} shows the optimized frequency placements, and Fig.~\ref{fig:base-fidelities} reports gate fidelities across module sizes. The geometric mean for gate fidelity for each module size:
\begin{itemize}
\item $N=2$: $1 - \epsilon_\text{gate} \approx 0.996$
\item $N=3$: $1 - \epsilon_\text{gate} \approx 0.994$
\item $N=4$: $1 - \epsilon_\text{gate} \approx 0.991$
\item $N=5$: $1 - \epsilon_\text{gate} \approx 0.940$
\end{itemize}
Fidelity degrades as expected with module size due to increasingly tight spectral constraints.

\section{Discussion}

\subsection{Reducing Connectivity to Improve Fidelity}

Full connectivity is not always necessary within a module. A module with $N$ qubits supports ${N \choose 2}$ interactions, but selectively omitting one or two gates can significantly improve overall fidelity. As shown in Fig.~\ref{fig:base-fidelities}, a 4-qubit module with one gate removed (5 active) improves average fidelity from $0.991$ to $0.993$, and with two gates removed (4 active) reaches $0.994$. These gains arise because omitting spectrally crowded interactions reduces pressure, enabling cleaner frequency separation for remaining gates. The optimizer reallocates detunings to prioritize fidelity across the active subset.

In NISQ devices, where fidelity is a limiting factor, our results suggest that selectively reducing connectivity may be preferable to enforcing full all-to-all coupling. However, realizing sparse physical layouts introduces practical challenges. Dedicated couplers for each interaction increase hardware complexity, chip area, and cryogenic routing demands. Instead, shared couplers—such as SNAILs—offer a more scalable alternative, reducing overhead while enabling compact designs and flexible control.

\subsection{Optimization Results and Design Considerations}

Frequency allocation for two-qubit gates in SNAIL-based architectures presents a nonconvex and highly variable optimization landscape. Individual runs of the optimizer often converge to distinct local minima, and high-fidelity solutions typically require hundreds of random restarts. Despite this variability, the procedure is practically sufficient: the goal is not to guarantee a global minimum but to find any configuration that satisfies design and fabrication requirements. While better solutions may exist, the marginal benefit of exhaustive search is outweighed by its computational cost.

Since gate durations are typically short relative to qubit lifetimes, incoherent errors contribute minimally. Figure~\ref{fig:fidelity_vs_terms} confirms that infidelity is dominated by coherent spectator effects when gate-mode detunings fall below roughly 200 MHz. Thus, fidelity improvements will likely come from suppressing coherent errors, through more precise frequency allocation, improved control, or reduced coupling strengths ($\eta$, $\lambda$, $g_3$), albeit at the cost of slower gates.

Table~\ref{tab:app-gate-params1} summarizes the gate fidelities for one such optimized frequency stack in a 4-qubit module. Achieving average gate fidelities above 0.99 required many hundreds of attempts, with typical solutions falling just below that threshold. In the reported case, average fidelities were $F_{\text{coh}} = 0.9947$ and $F_{\text{gate}} = 0.9918$, reaffirming that coherent errors remain the dominant limitation. Although the SNAIL-subharmonic penalty in our cost function is heuristic, it helps steer gate frequencies away from known unstable regions associated with SNAIL breakdown. While omitting this term does not drastically degrade results, including it ensures more robust solutions by avoiding poorly understood operational regimes.

\begin{table}[H]
\centering
\caption{Two-qubit gate parameters after frequency optimization. $F_{\text{coh}}$ reports coherent gate fidelity (excluding lifetime loss), and $F_{\text{gate}}$ includes incoherent decay.}
\label{tab:app-gate-params1}
\begin{tabular}{|c|c|c|c|}
\hline
\textbf{Gate} & \textbf{Freq (GHz)} & $\boldsymbol{F_{\text{coh}}}$ & $\boldsymbol{F_{\text{gate}}}$ \\
\hline
(Q0, Q1) & 0.2000 & 0.9980 & 0.9967 \\
(Q1, Q2) & 0.3200 & 0.9939 & 0.9926 \\
(Q0, Q2) & 0.5200 & 0.9864 & 0.9849 \\
(Q0, Q3) & 1.8797 & 0.9930 & 0.9894 \\
(Q1, Q3) & 2.0797 & 0.9958 & 0.9912 \\
(Q2, Q3) & 2.3997 & 0.9909 & 0.9870 \\
\hline
\end{tabular}
\end{table}

Because the optimization landscape is rugged, solutions are not unique. For example, some runs place two qubits above and two below the SNAIL frequency, while others put three above and one below. These variations allow some layout flexibility and can be selected to accommodate fabrication tolerances or simplify chip design. In more complex configurations with overlapping modules~\cite{yusuf2025densely}, additional frequency-sharing constraints must also be satisfied for qubits participating in multiple couplers—posing new optimization challenges that remain an active direction for future refinement. As overlapping complexity increases, it may become preferable to abandon continuous cost minimization in favor of discrete constraint satisfaction approaches.

Looking forward, we plan to explore noise-aware edge deletion and topological rewiring, including qubit-graph isomorphisms that preserve code distance while improving gate quality~\cite{nation2023suppressing}. These ideas align with recent architectural proposals emphasizing modular, fault-tolerant layouts~\cite{singh2024modular, sutcliffe2025distributed}.

Finally, direct hybridization between couplers remains a key limitation. In future designs, destructive interference via symmetric detuning may help suppress unwanted SNAIL–SNAIL interactions~\cite{Raman2018, majer2007coupling}. Increasing hybridization strength $\lambda$ or nonlinear coupling $g_3$ may enable faster gates, but must be balanced against increased crosstalk and instability. Mitigation strategies such as pulse shaping, compensatory drive cancellation, and error-aware scheduling could further reduce residual spectator terms~\cite{zajac2021spectator, seif2024suppressing, lu2023compilation, valles2025optimizing, theis2016simultaneous}. While our current model captures key hardware constraints, refined characterization of nonlinear dynamics and breakdown thresholds will be essential as next-generation devices emerge.

\section{Conclusion}
We have presented a method for optimizing frequency allocation in superconducting qubit modules that balances the competing effects of spectator interactions and $T_1$-limited incoherent loss. By modeling both coherent and incoherent infidelity contributions and incorporating empirical constraints, we produce frequency layouts that achieve high-fidelity gate sets even in spectrally dense modules.

To further improve performance, we demonstrated that selectively removing low-quality gates can raise overall fidelity with minimal impact on connectivity. These techniques offer a path forward for both near-term quantum devices and future fault-tolerant systems. Future work will incorporate additional nonidealities—such as AC Stark shifts, higher-order couplings, and advanced pulse synthesis—to support scalable, high-performance quantum architectures under realistic hardware constraints.

\section*{ACKNOWLEDGMENTS}
The data that support the findings of this article are openly available. Portions of this manuscript, including select figures and equations, are adapted from preliminary results presented at the 2nd StableQ Quantum System Stability and Reproducibility Workshop (StableQ'24)~\cite{mckinney2024towards}. The complete source code for this project is available at \url{https://github.com/PITT-HATLAB/corral_crowding}. This work is supported by the Army Research Office under Grants No. W911NF2310253. The views and conclusions contained in this document are those of the authors and should be interpreted as representing official policies, either expressed or implied, of the Army Research Office or the US Government. The US Government is authorized to reproduce and distribute reprints for government purposes notwithstanding any copyright notation herein.

\bibliography{0_refs}

\end{document}